\documentclass[a4paper]{jpconf}
\usepackage{graphicx}
\begin{document}
\title{Study of azimuthal correlations between D mesons and charged particles with the ALICE experiment}

\author{Fabio Colamaria, on behalf of the ALICE Collaboration}

\address{University of Bari ``Aldo Moro'' and INFN - Section of Bari}

\ead{fabio.colamaria@ba.infn.it}

\begin{abstract}
A study of azimuthal correlations between D$^0$, D$^+$, and D$^{*+}$ and charged particles in pp collisions at $\sqrt{s}$ = 7 TeV and p--Pb collisions at $\sqrt{s_{\rm NN}}$ = 5.02 TeV is presented. D mesons were reconstructed from their hadronic decays at central rapidity ($|y| < 0.5$) in the transverse momentum range $3 \leq p_{\rm T}^{\rm D} \leq 16$ GeV/$c$ and they were correlated with charged particles reconstructed in the pseudorapidity range $|\eta| < 0.8$. Perspectives for the measurement in Pb--Pb collisions at $\sqrt{s_{\rm NN}}$ = 5.5 TeV after the ALICE upgrade are presented as well.
\end{abstract}

\section{Physics motivations}
\label{sec:intro}
ALICE~\cite{bib:ALICE} has measured $p_{\rm T}$-differential cross sections for D-meson production at central rapidity in pp~\cite{bib:Dmes_pp_7, bib:Ds_pp_7, bib:Dmes_pp_2.76}, p--Pb~\cite{bib:Dmes_pPb} and Pb--Pb~\cite{bib:Dmes_PbPb} collisions. A suppression of the D-meson yield by a factor 4$-$5 was observed in central Pb--Pb collisions for $p_{\rm T} >$ 5 GeV/$c$ with respect to cross sections measured in pp collisions, scaled by the nuclear overlap function. This feature can be attributed to the interaction of heavy quarks with the Quark-Gluon plasma formed in such collisions. The analysis of angular correlations between D mesons and unidentified charged particles in Pb--Pb collisions can provide further insight into this topic, allowing us to address medium-induced modifications of charm fragmentation and hadronization and the dependence of the energy loss on the path length the parton travels through the medium.

Such a study is of great importance also in pp and p--Pb collisions. In pp collisions it allows us to characterize charm-quark jets and study their properties, providing at the same time a reference for the measurements in p--Pb and Pb--Pb collisions. In p--Pb collisions it enables the study of cold nuclear matter effects and the search, in the heavy-flavour sector, for long-range ridge-like structures (``double ridge'') in near and away-side regions, already observed in hadron-hadron correlations~\cite{bib:DoubleRidge}.

\section{Description of the analysis}
\label{sec:descr}
The analysis was carried out for D$^0$, D$^+$ and D$^{*+}$ mesons on minimum bias samples of $3.1 \cdot 10^8$ pp collisions at $\sqrt{s} = 7$ TeV and $1.0 \cdot 10^8$ p--Pb collisions at $\sqrt{s_{\rm NN}} = 5.02$ TeV, for different ranges of the D-meson $p_{\rm T}$ and different thresholds on the charged particle $p_{\rm T}$.
D-meson candidates (\textit{trigger} particles) were reconstructed from their hadronic decay channels (D$^0\rightarrow$ K$^-\pi^+$, D$^+\rightarrow$ K$^-\pi^+\pi^+$ and D$^{*+}\rightarrow$ D$^0 \pi^+\rightarrow$ K$^-\pi^+\pi^+$) and selected exploiting their displaced decay vertex topology, PID (particle identification) and reconstruction quality cuts on the daughter tracks. The values of $S/B$ are in the range 0.25 to 2 (up to 5.6 for the D$^{*+}$), depending on the D-meson $p_{\rm T}$. The selected D-meson candidates in the signal region ($\pm 2\sigma$ from the centre of the signal peak in the invariant mass distribution) were then correlated with charged particles (\textit{associated tracks}) within $|\eta| < 0.8$ and satisfying a selection on their reconstruction quality. The difference in pseudorapidity $\Delta\eta$ and azimuthal angle $\Delta\varphi$ between the D-meson candidate and the charged-particle track was evaluated.

The contribution of background candidates was subtracted using the correlation distribution from the candidates in the sidebands of the invariant mass distribution (the regions 4$-$8$\sigma$ away from the signal peak), normalized by the ratio of background candidates in signal and sideband regions. Several corrections were applied to the correlation distributions to account for: (i) \textit{Associated track reconstruction efficiency}: each D--hadron pair was weighted by the inverse of the track reconstruction efficiency of the associated track, evaluated from simulations; (ii) \textit{D-meson selection efficiency}: each correlation entry was weighted by the inverse of the D-meson reconstruction and selection efficiency, computed from simulations; (iii) \textit{Limited detector acceptance and detector spatial inhomogeneities}: a correction factor was obtained exploiting the event mixing technique, i.e.~correlating D-meson candidates from one event with tracks from other events with similar multiplicity and position of the primary vertex in the beam direction; (iv) \textit{Beauty feed-down}: MC simulations based on PYTHIA were exploited to evaluate a template distribution of angular correlations between D mesons from B-meson decays and charged particles; this distribution was scaled to match the the proper ratio of feed-down to prompt D mesons, calculated as described in~\cite{bib:Dmes_pp_7}, and subtracted from the inclusive D--hadron correlation distribution obtained from data; (v) \textit{Secondary track contamination}, i.e.~tracks from strange-hadron decays or produced in interactions of particles with the detector material. From a MC study, no visible angular dependence was found for this contamination. A global scale factor, corresponding to the fraction of primary particles in the sample of associated tracks, was thus applied to the correlation distributions.

Due to the limited statistics available, the fully corrected 2D correlations were projected onto the $\Delta\varphi$ axis, producing azimuthal correlation distributions, which were normalized to the number of trigger D mesons. A weighted average of the results for the three meson species, consistent among themselves within the uncertainties, was then performed.
A fit function, composed of a constant term plus two Gaussian functions with a periodicity condition, was applied to the azimuthal correlation distributions.
The two Gaussian functions, centered at $\Delta\varphi \sim 0$ and $\Delta\varphi \sim \pi$, model the near and away-side peaks in the correlation distribution.
Physical observables (near-side yield, near-side peak width, height of the baseline of the distribution, defined as the average height of the points in the transverse region, far from the peaks) were extracted and compared between the two collision systems and with predictions from Monte Carlo simulations. Finally, a careful evaluation of all the sources of systematic uncertainties was performed.

\section{Results in pp and p--Pb collision systems}
\label{sec:results}
\begin{figure}[!t]
\centering
\begin{minipage}{\linewidth}
  \centering
  $\vcenter{\hbox{\includegraphics[width=.41\linewidth]{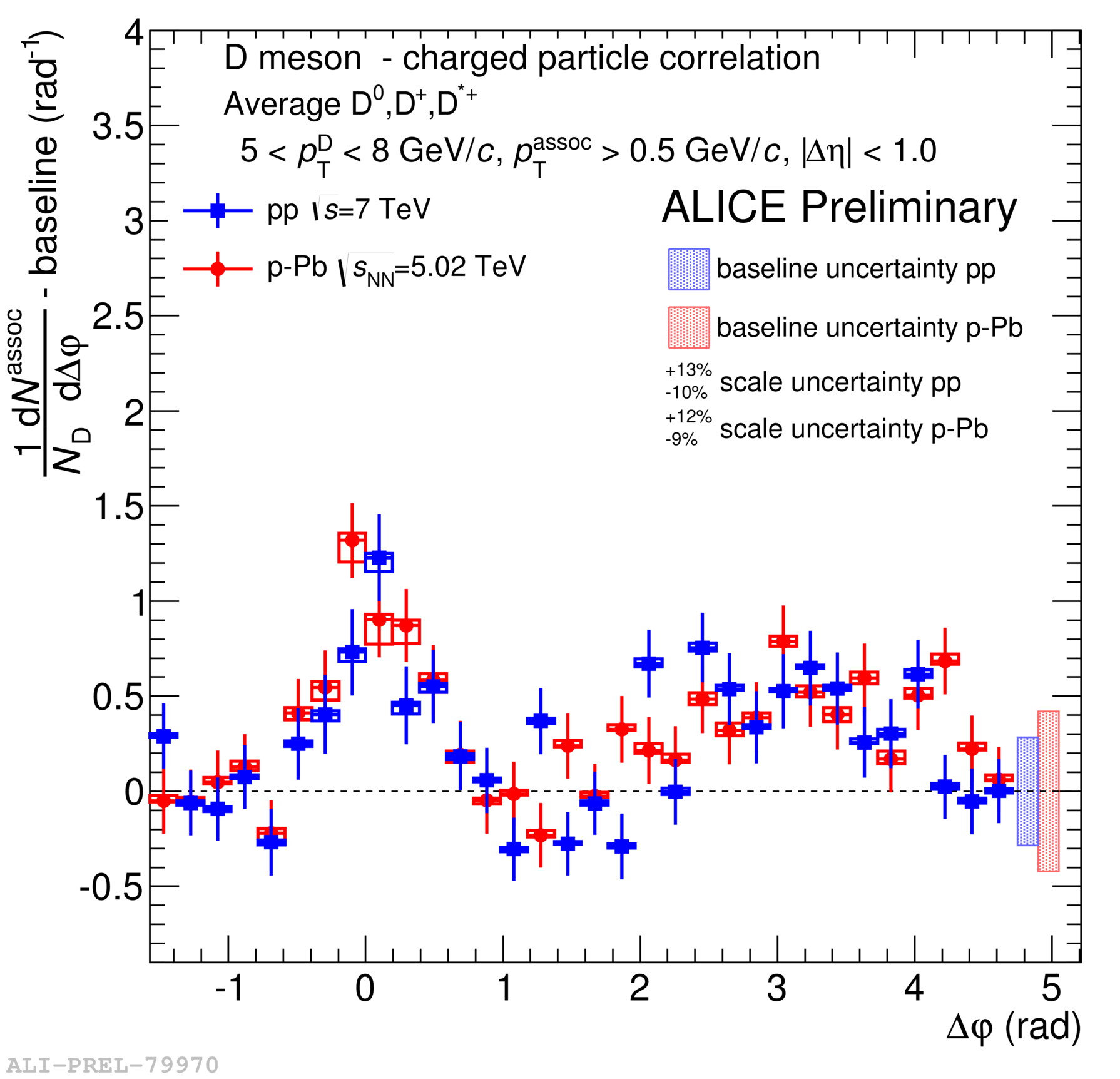}}}$ \hspace{0.5cm}
  $\vcenter{\hbox{\includegraphics[width=.41\linewidth]{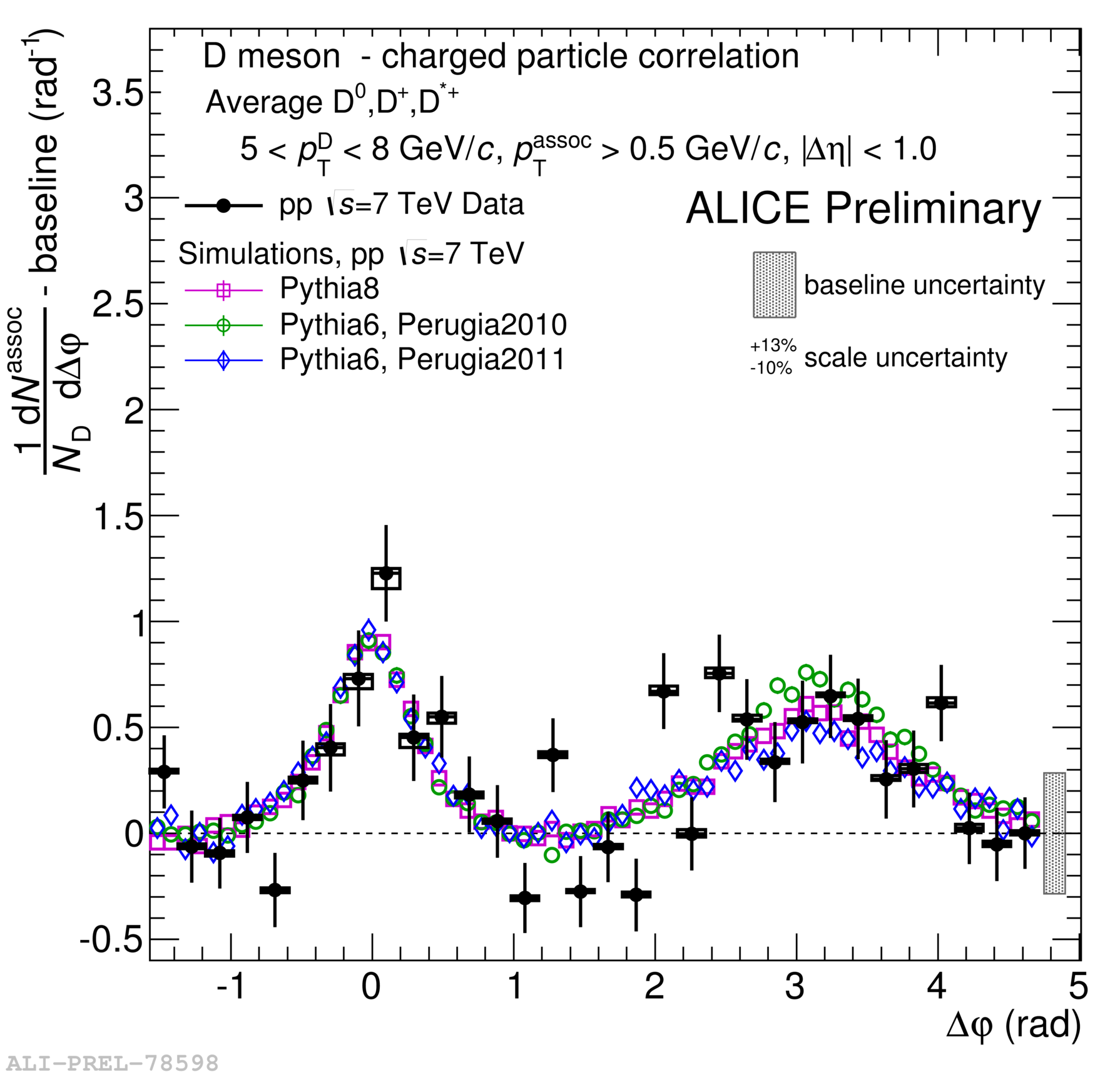}}}$
\end{minipage}
\caption{Left: comparison of D--hadron azimuthal correlation distributions in pp and p--Pb collisions, after baseline subtraction, from the weighted average of D$^0$, D$^+$ and D$^{\ast +}$ measurements. Right: comparison of D--hadron azimuthal correlation distribution and predictions from different tunes of the PYTHIA generator, for pp collisions, after baseline subtraction. In both panels, statistical and uncorrelated systematic uncertainties are shown as error bars and boxes, respectively.}
\label{fig:RisData}
\end{figure}

\begin{figure}[!h]
\centering
\begin{minipage}{\linewidth}
  \centering
  $\vcenter{\hbox{\includegraphics[width=.41\linewidth]{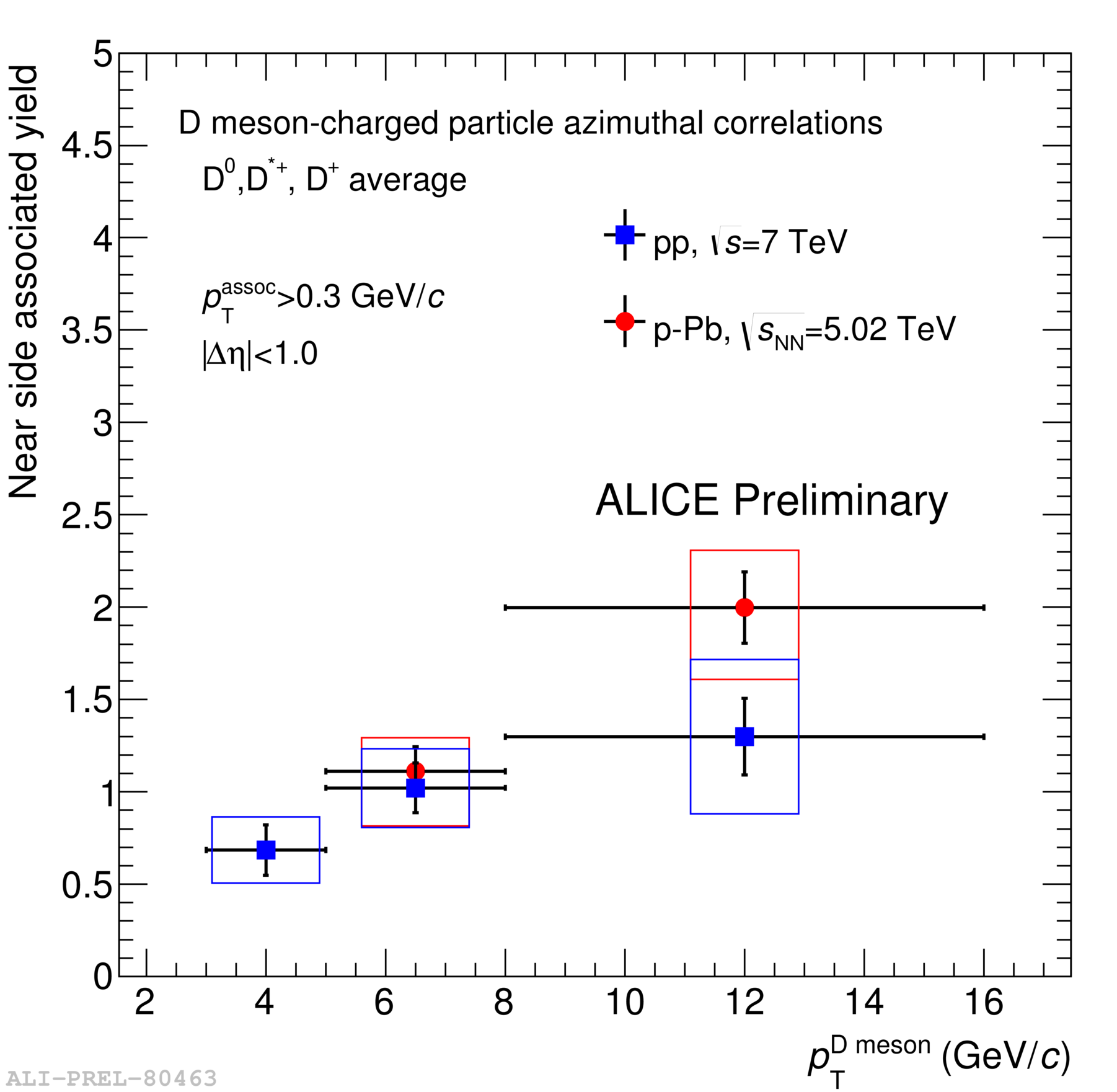}}}$ \hspace{0.5cm}
  $\vcenter{\hbox{\includegraphics[width=.41\linewidth]{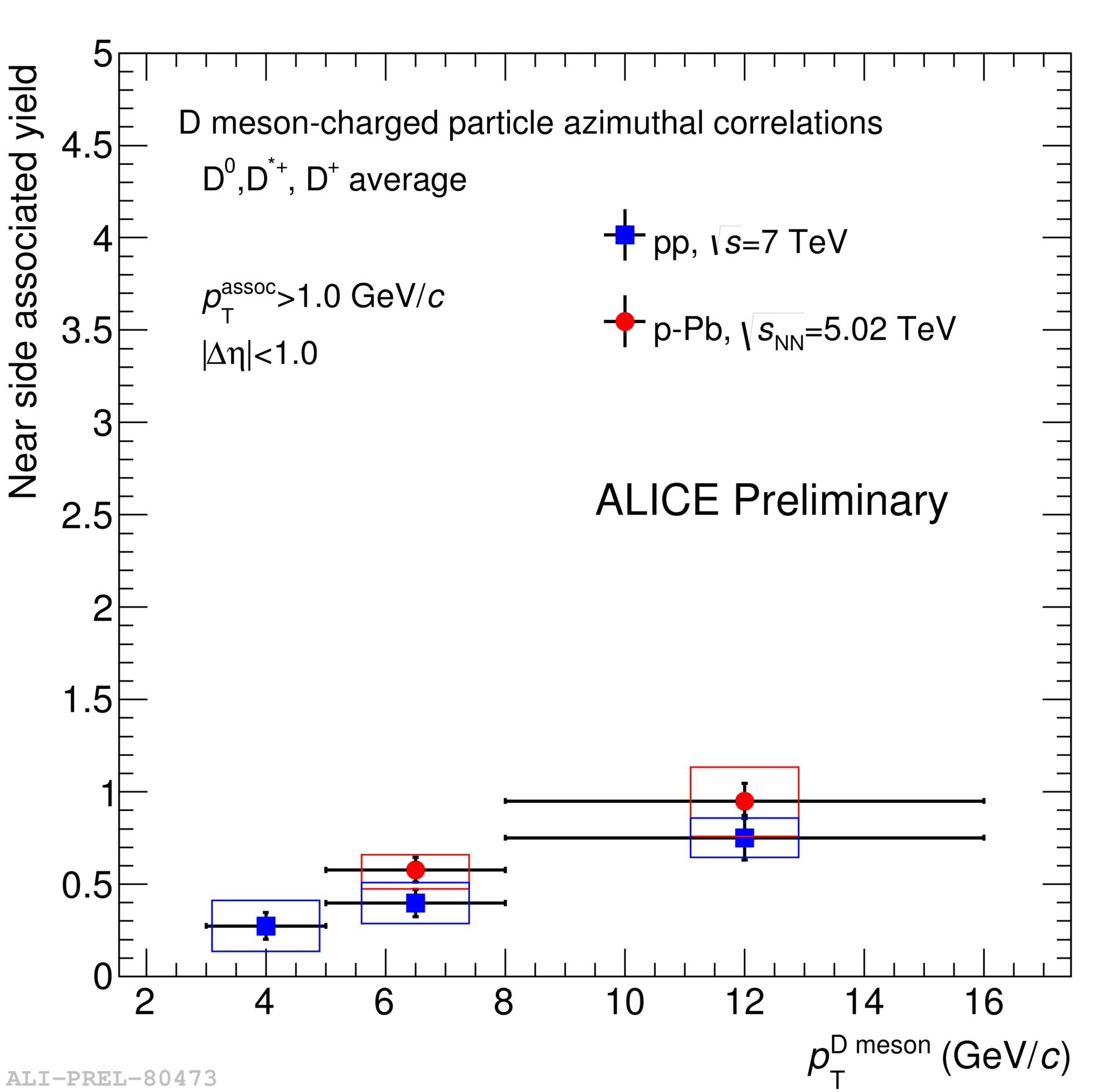}}}$
\end{minipage}
\caption{Comparison of near-side associated yields extracted in pp and p--Pb collisions as a function of the D-meson $p_{\rm T}$ for $p_{\rm T}^{\rm assoc} > 0.3$ (left panel) and 1 (right panel) GeV/$c$. Statistical and uncorrelated systematic uncertainties are shown as error bars and boxes, respectively.}
\label{fig:NSY}       
\end{figure}

The left panel of Fig.~\ref{fig:RisData} shows a comparison of D--hadron correlations in pp and p--Pb collisions, after the subtraction of the baseline, for $5< p_{\rm T}^{\rm D} < 8$ GeV/$c$ and associated tracks with $p_{\rm T}^{\rm assoc} > 0.5$ GeV/$c$. In both systems, the near and away-side peaks are clearly visible, despite the statistical fluctuations, and the correlation pattern is similar for pp and p--Pb collisions. In the right panel of Fig.~\ref{fig:RisData} the baseline-subtracted D--hadron correlation distribution from the pp analysis is compared with the outcome of PYTHIA simulations, at $\sqrt{s} = 7$ TeV, exploiting different tunes of the simulation parameters. The PYTHIA results with all three parameter tunes provide a fair description of the measured distribution within uncertainties.
A more quantitative comparison between pp and p--Pb results for the near-side region can be performed by extracting the per-trigger yield in the near-side peaks from the fit function applied to the correlation distributions. The near-side associated yields obtained from pp and p--Pb correlation distributions are shown in Fig.~\ref{fig:NSY} as a function of the D-meson $p_{\rm T}$ for associated tracks with $p_{\rm T}^{\rm assoc} > 0.3$ GeV/$c$ (left) and $p_{\rm T}^{\rm assoc} > 1.0$ GeV/$c$ (right). The pp and p--Pb yields are compatible within the total uncertainties. Given the size of the current uncertainties, no firm conclusions on possible modifications of D--hadron correlations due to cold nuclear matter effects can be drawn.

\section{Perspectives for Pb--Pb collisions}
\label{sec:lead}
The D--hadron correlation measurement is not feasible on the currently available sample of Pb--Pb collisions, with the current performance of the ALICE detectors. The very low D-meson signal/background ratio and the large number of tracks from the underlying event, not correlated with D mesons, induce substantial statistical fluctuations in the $\Delta\varphi$ distributions when the background is subtracted. This washes out any correlation structure possibly present.

\begin{figure}[!h]
\hspace{2.5pc}
\includegraphics[width=.40\linewidth]{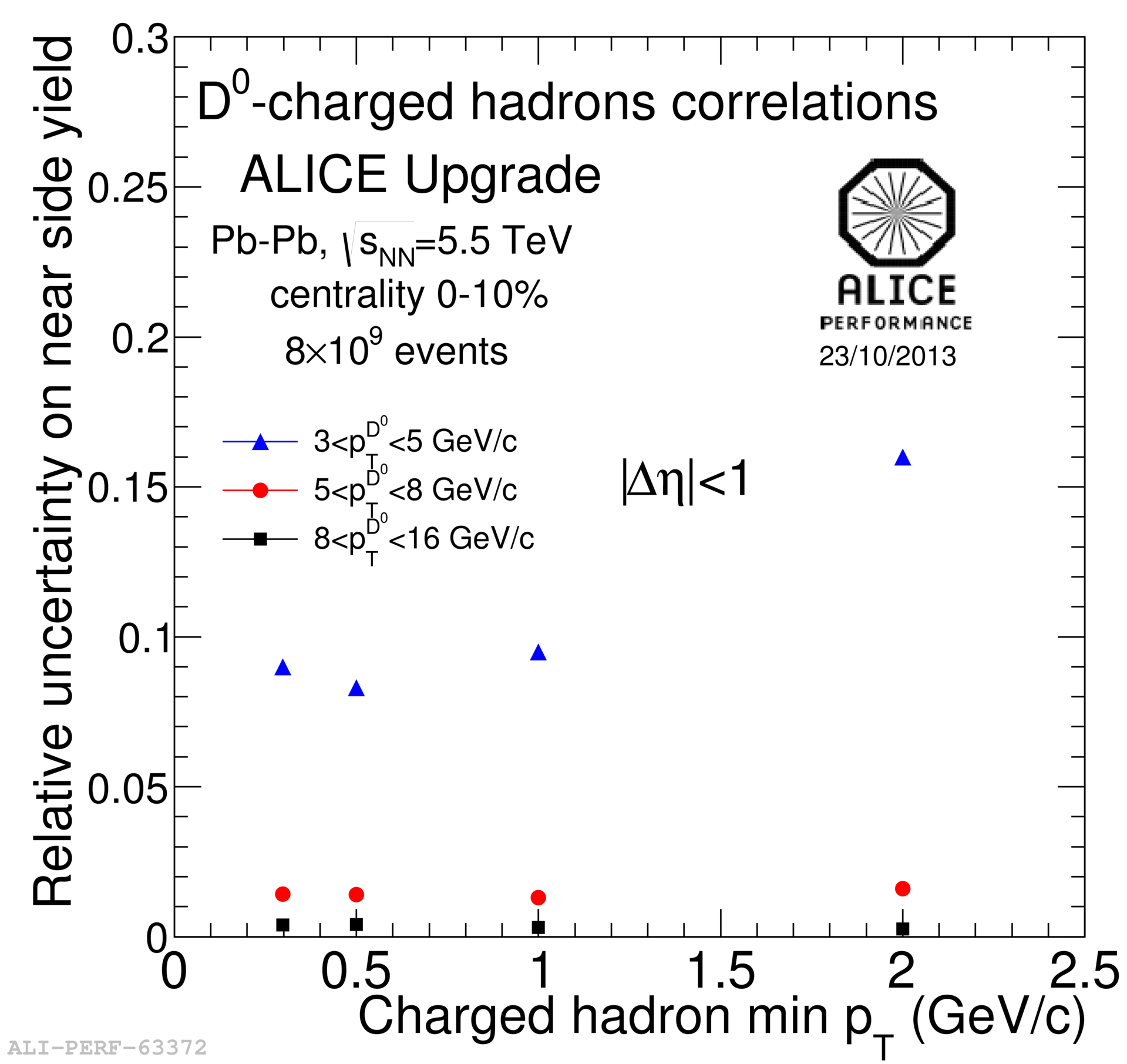}\hspace{2pc}%
\begin{minipage}[b]{18pc}\caption{\label{fig:pbpbUpg}Estimates for the statistical uncertainty on the near-side yield from the analysis of the expected Pb--Pb statistics after the ALICE upgrade, for different kinematic ranges, obtained for an integrated luminosity of $L_{\rm int} = 10$ nb$^{-1}$.}
\end{minipage}
\end{figure}

After the ALICE upgrade (expected for 2018-2019)~\cite{bib:LoI, bib:UpgITS}, though, a dramatic improvement of tracking and vertexing performance is expected, which will lead to an increase of the $S/B$ ratio by a factor up to 10 for the D-meson reconstruction. Combined with the increase of a factor $\approx$100 in the ALICE readout rate for minimum-bias Pb--Pb collisions, this will enable the study of D--hadron correlations also in this system.
A simulation of the analysis performance on central (0-10\%) Pb--Pb collisions was carried out using a template of correlation distributions from PYTHIA. As shown in Fig.~\ref{fig:pbpbUpg}, the results of this study indicate a very low statistical uncertainty on the evaluation of the near-side yield, below 1\% for $p_{\rm T}^{{\rm D}^0} > 8$ GeV/$c$ and about 10\% for $3 < p_{\rm T}^{{\rm D}^0} < 5$ GeV/$c$, for an integrated luminosity of $L_{\rm int} = 10$ nb$^{-1}$. A substantial reduction of the uncertainties is expected also for pp and p--Pb collision systems. After the ALICE upgrade, thus, it should be possible to compare the results in the three collision systems and to quantify the effects of the charm quark energy loss in a QGP medium.
\section{Conclusions}
\label{sec:lead}
ALICE measurements of D--hadron angular correlations in pp and p--Pb collisions are compatible, within uncertainties, in terms of the correlation shapes and the near-side associated yields. Consequently, no evident influence of cold nuclear matter on the charm fragmentation and hadronization can be claimed. The correlation distributions are well described by the outcome of PYTHIA simulations as well.
While the analysis cannot be currently performed in Pb--Pb collisions, it will become feasible also in this collision system after the ALICE upgrade (2018-2019), when the tracking and vertexing performance of the experiment will be largely enhanced.

\end{document}